\documentstyle[times, epsfig]{mn}

\newif\ifAMStwofonts

\ifoldfss
  \ifCUPmtlplainloaded \else
    \NewTextAlphabet{textbfit} {cmbxti10} {}
    \NewTextAlphabet{textbfss} {cmssbx10} {}
    \NewMathAlphabet{mathbfit} {cmbxti10} {} 
    \NewMathAlphabet{mathbfss} {cmssbx10} {} 
  \fi
  \ifAMStwofonts
    \ifCUPmtlplainloaded \else
      \NewSymbolFont{upmath} {eurm10}
      \NewSymbolFont{AMSa} {msam10}
      \NewMathSymbol{\upi}     {0}{upmath}{19}
      \NewMathSymbol{\umu}     {0}{upmath}{16}
      \NewMathSymbol{\upartial}{0}{upmath}{40}
      \NewMathSymbol{\leqslant}{3}{AMSa}{36}
      \NewMathSymbol{\geqslant}{3}{AMSa}{3E}

       \let\le=\leqslant
       
    \fi
  \fi
\fi 

\ifnfssone
  \newmathalphabet{\mathit}
  \addtoversion{normal}{\mathit}{cmr}{m}{it}
  \addtoversion{bold}{\mathit}{cmr}{bx}{it}
  \newmathalphabet{\mathbfit} 
  \addtoversion{normal}{\mathbfit}{cmr}{bx}{it}
  \addtoversion{bold}{\mathbfit}{cmr}{bx}{it}
  \newmathalphabet{\mathbfss} 
  \addtoversion{normal}{\mathbfss}{cmss}{bx}{n}
  \addtoversion{bold}{\mathbfss}{cmss}{bx}{n}
  \ifAMStwofonts
    \ifCUPmtlplainloaded \else
and
your
      \UseAMStwoboldmath
      \makeatletter
      \new@mathgroup\upmath@group
      \define@mathgroup\mv@normal\upmath@group{eur}{m}{n}
      \define@mathgroup\mv@bold\upmath@group{eur}{b}{n}
      \edef\UPM{\hexnumber\upmath@group}
      \new@mathgroup\amsa@group
      \define@mathgroup\mv@normal\amsa@group{msa}{m}{n}
      \define@mathgroup\mv@bold\amsa@group{msa}{m}{n}
      \edef\AMSa{\hexnumber\amsa@group}
      \makeatother
      \mathchardef\upi="0\UPM19
      \mathchardef\umu="0\UPM16
      \mathchardef\upartial="0\UPM40
      \mathchardef\leqslant="3\AMSa36
      \mathchardef\geqslant="3\AMSa3E

       \let\le=\leqslant

    \fi
  \fi
\fi 

\ifnfsstwo
  \DeclareMathAlphabet{\mathbfit}{OT1}{cmr}{bx}{it}
  \SetMathAlphabet\mathbfit{bold}{OT1}{cmr}{bx}{it}
  \DeclareMathAlphabet{\mathbfss}{OT1}{cmss}{bx}{n}
  \SetMathAlphabet\mathbfss{bold}{OT1}{cmss}{bx}{n}
  \ifAMStwofonts
    \ifCUPmtlplainloaded \else
      \DeclareSymbolFont{UPM}{U}{eur}{m}{n}
      \SetSymbolFont{UPM}{bold}{U}{eur}{b}{n}
      \DeclareSymbolFont{AMSa}{U}{msa}{m}{n}
      \DeclareMathSymbol{\upi}{0}{UPM}{"19}
      \DeclareMathSymbol{\umu}{0}{UPM}{"16}
      \DeclareMathSymbol{\upartial}{0}{UPM}{"40}
      \DeclareMathSymbol{\leqslant}{3}{AMSa}{"36}
      \DeclareMathSymbol{\geqslant}{3}{AMSa}{"3E}

       \let\le=\leqslant

    \fi
  \fi
\fi 

\ifCUPmtlplainloaded \else
  \ifAMStwofonts \else 
    \def\upi{\pi}
    \def\umu{\mu}
    \def\upartial{\partial}
  \fi
\fi

\title{On the evolution of the Fe abundance and of the Type Ia SN rate in clusters of galaxies}

\author[F. Calura, F. Matteucci, P. Tozzi]
       {F. Calura$^{1}$\thanks{E-mail: fcalura@ts.astro.it}, 
        F. Matteucci$^{1,2}$, P. Tozzi$^{1,3}$\\
	(1) INAF - Osservatorio Astronomico di Trieste, via G.B. Tiepolo 11, 34131 Trieste, Italy \\
        (2) Dipartimento di Astronomia-Universit\'a di Trieste, Via G.B. Tiepolo
	11, 34131 Trieste, Italy\\
	(3) INFN, Sezione di Trieste, Via Valerio 2, 34127, Trieste, Italy \\
	 }
	
\date{}

\pagerange{\pageref{firstpage}--\pageref{lastpage}}
\pubyear{2007}

\begin{document}

\maketitle

\label{firstpage}
\begin{abstract}

The study of the Fe abundance in the intra cluster medium (ICM) 
provides 
strong constraints on the integrated 
star formation history and supernova rate of the cluster galaxies, as well as on the ICM enrichment mechanisms. 
In this Letter, using chemical evolution models for galaxies of different morphological types, we study the 
evolution of the Fe content of clusters of galaxies.  We assume that the ICM Fe enrichment occurs by means of galactic winds 
arising from elliptical galaxies and from gas stripped from the progenitors of S0 galaxies via external mechanisms, due to the 
interaction of the inter stellar medium with the ICM. 
The Fe-rich gas ejected by ellipticals accounts for the $X_{Fe, ICM}$ values observed at $z > 0.5$, whereas the gas stripped from the progenitors of 
the S0 galaxies accounts for the increase of $X_{Fe, ICM}$ observed at $z<0.5$. 
We tested two different scenarios for Type Ia supernova (SN) progenitors  
and we model the Type Ia SN rate observed in clusters, finding a good agreement between our predictions and the available observations. 
\end{abstract} 


\section{Introduction}
The hot intracluster gas contains a large amount of highly-ionized heavy elements, 
which can be detected by means of X-ray observations. Among these heavy elements, Fe is the easiest to detect in the X-rays and 
its abundance in the ICM provides 
strong constraints on the integrated 
star formation history and supernova rate (SNR) of the cluster galaxies (Matteucci \& Vettolani 1988, Renzini 1997). 
In the last few years, thanks to the deep X-ray observations with the
Chandra and XMM satellites, 
it has been possible to study the evolution of the 
intracluster Fe abundance 
out to redshift $z \sim 1.3$, corresponding to a lookback time of $\sim 9$ Gyr. 
Tozzi et al. (2003), found that the average Fe abundance of the ICM
at $z \sim 1$ is comparable with the local value $X_{Fe}
\sim 0.45 X_{Fe,\odot}$ \footnote{For the solar Fe mass fraction $X_{Fe,\odot}$, by using the meteoritic Fe
abundance in mass by Anders \& Grevesse (1989) and the H mass fraction
$X_{H}=0.75$ (Lodders 2003), we obtain a value of $X_{Fe,\odot}=1.3
\cdot 10^{-3}$. This value is in  
agreement with the most recent determination of the solar photometric Fe abundance by Asplund et al. (2005). 
Note that most of the other papers on clusters use instead the photospheric Fe abundance of Anders \& Grevesse (1989).}, 
without detecting significant evidence of
evolution from a sample of 18 clusters at $0.3<z<1.2$.  
In a more recent paper, Balestra et al. (2007) extended the analysis of 
Tozzi et al. (2003) to a sample of 56 clusters with redshifts $0.3 \le z \le 1.3$. 
For $z>0.5$ they found a constant average Fe abundance 
$X_{Fe} \sim 0.38 X_{Fe,\odot}$, whereas for  $z<0.5$ they found  significant evidence of evolution, parametrized by 
a power law $X_{Fe}(z) \propto (1+z)^{-1.25}$ and implying that at the present time, the average 
ICM Fe abundance is a factor of 2 larger than at $z\sim 0.5$ in the central regions, at radii $R \le 0.3  R_{vir}$.\\
The main producers of Fe in clusters are Type Ia supernovae (SNe)  
(Matteucci \&  Vettolani 1988, Renzini et al. 1993, Pipino et al. 2002, Ettori 2005, Loewenstein 2006).
Two possible mechanisms have been proposed to explain the ICM Fe enrichment, 
i.e. SN-driven galactic winds (Renzini et al. 1993, Pipino et al. 2002) 
and enrichment processes linked to environmental effects,
such as ram pressure stripping, tidal stripping or viscous stripping of Fe-rich gas from cluster 
galaxies (Gunn \& Gott 1972, Domainko et al. 2006, see also Cora et al., in prep.).  
In this Letter, by means of chemical evolution models for galaxies of different morphological types, we study the 
evolution of the Fe content of clusters of galaxies.  
We start from two main observables, i.e. the Fe abundance and the Type Ia SN rate, and test different  models 
which can account for the observed evolution of both quantities. 
Our aim is to determine which galaxies are the main responsible for 
the observed evolution 
of the Fe content of the ICM, and what are the relative roles of galactic winds and environmental effects in this process. 
The plan of this Letter is as follows. In section 2, we describe the chemical evolution models. 
In section, 3 we present our results and their comparison with the observed abundances. Finally, in section 4, some conclusions 
are drawn.

\section{The chemical evolution models}
In this Letter, we use two chemical evolution models, one describing an elliptical galaxy, 
the other describing an S0 galaxy. We assume that these two kinds of objects 
are responsible for the chemical enrichment of the ICM. 
This is supported by the fact that in local clusters, 
the intra-cluster Fe mass is correlated only with the luminosity of elliptical and S0 (Arnaud et al. 1992). 
The common features between these two models are (i) the stellar nucleosynthesis prescriptions and (ii) the adopted stellar IMF. \\
(i) For massive stars, we adopt the Fe yield of Woosley \& Weaver (1995) for a solar chemical composition, 
which has proven to be the best choice for the Milky Way (Franc\c ois et a at. 2004). 
For the Type Ia SNe, we adopt the Fe yields of Iwamoto et al. (1999). \\
(ii) For both models, we assume a Salpeter (1955) IMF constant in time. \\
A detailed description of the chemical evolution equations used in this work can be found in 
Matteucci \& Greggio (1986). 
For the chemical evolution of ellipticals we adopt the model of Pipino \& al. (2002),  
where we address the reader for details. 
The elliptical galaxy forms by means of a rapid collapse of pristine gas where star 
formation occurs at a very high rate ($\sim 1000 M_{\odot} yr^{-1}$). After a timescale which depends on its mass,   
the galaxy develops a galactic wind 
due to the energy deposited by SNe into the interstellar medium (ISM). After 
the onset of the wind, no star formation (SF) is assumed to take place.
The winds develops as soon as the thermal energy of the ISM exceeds its gravitational binding 
energy. The thermal energy is evaluated by taking into account feedback from SNe and stellar winds. 
The binding energy is calculated considering a 
massive, diffuse dark matter halo  around the galaxy.  
The main parameters of this model are 
the baryonic mass $M_{lum}=10^{11} M_{\odot}$, the SF efficiency, $\nu_{E} =11 Gyr^{-1}$ and an effective radius 
$R_{eff}=3$ kpc (see Table 1 of Pipino et al. 2002).\\ 
We assume that S0 galaxies originate from disc galaxies, in which at a late stage 
SF stops owing to massive gas loss by means of some external mechanism, represented by 
interactions with the ICM, collisions or tidal stripping (Larson et al. 1980, Boselli \& Gavazzi 2006). 
We assume that, as soon as the SF stops, the galaxy starts to lose enriched gas at a constant rate. 
At the final stage of its evolution, the S0 galaxy has ejected all of its gas and consists only of a stellar disc.\\ 
This scenario is supported by a large set of observations. The first is the Butcher \& Hoemler (1978) 
effect, i.e. the increase with redshift of the fraction of blue cluster galaxies, which seems strictly connected 
to the decrease of the fraction of cluster S0 galaxies with redshift 
(Dressler 1997, Poggianti et al. 1999, van Dokkum et al. 2000, Fasano et al. 2000, Tran et al. 2005). Another indication comes from the presence 
of anemic spirals 
in clusters, which are likely to represent the missing link between cluster gas-rich discs and S0 galaxies 
(van den Bergh 1976, Elmegreen et al. 2002, Boselli et al. 2006). 
The fact that the S0 fraction is maximum in the cluster cores 
(Dressler et al. 1997, Helsdon \& Ponman 2003) tells us that environmental effects, rather than internal effects, 
are the main drivers for the formation of S0s.\\ 
To calculate the enrichment due to the gas loss accompanying the transition 
from gas-rich discs to S0s, we use the two infall model  
by Chiappini et al. (1997). In this model, the first infall creates the halo and the thick 
disc, on a timescale of $\sim 1$ Gyr.  The second infall gives rise to the thin disc. 
The spiral disc is approximated by several independent rings, 2 kpc wide, 
without exchange of matter between them. 
The timescale for the disc formation is assumed to increase 
with the galactocentric distance ($\tau= 8$ Gyr at the solar circle), 
thus producing an ``inside-out'' scenario for the disc formation. 
The main parameters of this model are 
the timescale for the inner halo formation ($\tau_H=$1 Gyr), 
the thin disc timescale $\tau_D$, which is a function of the galactic 
radius, and 
the total surface mass density profile at the present time, $\sigma_D(R)$, 
as a function of the radius $R$, 
for which we use an exponential law $\sigma_D(R) \propto e^{-R/R_{D}}$, 
 where $R_{D}=3.5$ kpc (Romano et al. 2000), in analogy with the Milky Way. 
The last parameter is  the SF efficiency 
$\nu_{S0} = 1 Gyr^{-1}$. 
In the disc, the time at which the SF is assumed to stop is determined in order to reproduce the observed 
Fe abundance in the ICM (section 3.1). With this choice of the parameters, for the disc of an average S0 galaxy we predict a 
present day stellar mass of $M_{*,S0} = 4 \cdot 10^{9} M_{\odot}$. \\
We adopt a Lambda-cold dark matter cosmology 
($\Lambda$CDM, $\Omega_{0}=0.3$, 
 $\Omega_{\Lambda}=0.7$) and $h=0.7$. 
For all galaxies, we assume that the star formation started at redshift $z_{f}=10$, corresponding to a lookback time of $13$ Gyr. 
The spectral evolution of all the galactic morphological types has been calculated by means 
of the spectro-photometric code by Jimenez et al. (2004). By means of this code, 
we have computed the present day colours, finding for S0s (B-V)=0.96-0.98 and (U-B)=0.47-0.5, in excellent 
agreement with the available observational values (Schweizer \& Seitzer 1992).

\subsection{The Type Ia SN rate}
We assume that Type Ia SNe originate from 
the explosion of a C/O white dwarf (WD) in a close binary system, 
where the companion is either a Red Giant or a Main-Sequence star. This is the  
single-degenerate (SD) model (Whelan \& Iben 1973). 
Here, for the Type Ia SN rate, we study the two different formulations discussed in Matteucci et al. (2006, hereinafter M06).  
In the first formulation, (Greggio \& Renzini 1983, Matteucci \& Recchi 2001, hereinafter MR01), the Type Ia SN rate is given by: 
\begin{equation}
 R_{Ia}(t) = A_{Ia} \int_{M_{Bm}}^{M_{BM}} \phi(M_{B}) [\int_{\mu_{m}}^{0.5} f(\mu)
\psi(t-\tau_{M_{2}})d\mu]dM_{B}
\end{equation}
where $A_{Ia}$ represents the fraction of binary systems 
with total mass  $M_{Bm}\le M_{B}/M_{\odot} \le M_{BM} $ which produce Type Ia SNe. 
The quantity $\mu=M_{2}/M_{B}$ is the ratio between the secondary component of the binary system (i.e. the originally less
massive one) and the total mass of the system $M_{B}$, and $f(\mu)$ is the 
distribution function of this ratio. Statistical studies 
(see MR01) indicate that mass 
ratios close to $0.5$ are preferred, so the formula:\\
\begin{equation}
 f(\mu)=2^{1+\beta}(1+\beta)\mu^{\beta}
\end{equation}
is commonly adopted, with $\beta=2$ (Matteucci \& Recchi 2001).
$\tau_{M_{2}}$ is the lifetime of the secondary star 
in the system, 
which determines the timescale for the explosion.
The assumed value of $A_{Ia}$ is fixed by reproducing the present time observed rate (Matteucci et al. 2003, M06). 
For the masses ${M_{Bm}}$ and ${M_{BM}}$, we
use the values $3M_{\odot}$ and $16M_{\odot}$, respectively (Greggio \& Renzini 1983).\\ 
The second formulation for the Type Ia SN rate is based on the recent results 
by Mannucci, Della Valle \& Panagia (2006, hereafter MVP06) which, on the basis of 
the relation between the Type Ia SN rate 
and the colour of the parent galaxies, 
their radio power as measured by Della Valle et al. (2005) and cosmic age, 
concluded that there are two populations of progenitors of Type Ia SNe. 
According to these results, the above observations can be accounted for 
if half of the SNe Ia explode within 10$^8$ yr after the formation 
of their progenitors, while the rest explode during 
a wide period of time extending up to 10 Gyr. 
Following Greggio (2005), 
in this new formulation, dubbed as MVP06, the Type Ia SN rate is: 
\begin{equation}
R_{Ia}(t)=k_{\alpha} \int^{min(t, \tau_x)}_{\tau_i}{ \psi(t-\tau) 
DTD(\tau) d \tau}, 
\end{equation}
where $k_{\alpha}$ is the number of stars per unit mass in a stellar 
generation and contains the IMF. The function $DTD(\tau)$ is the 
delay time distribution given in equations 7 and 8 of M06. 
A more detailed description of the two different formulations used 
to model the Type Ia SN rate, as well as their effect on the chemical 
evolution of spirals, ellipticals and irregular galaxies, 
can be found in M06.

\section{Results}
\subsection{The evolution of the Fe abundance in the ICM}

We assume that, immediately after the Fe ejection into the ICM, the Fe mass is  distributed homogeneously, i.e. 
we assume that the ICM is well mixed. 
We estimate the ICM abundance by evaluating, at each redshift $z$,  
the comoving density of the Fe ejected  
by ellipticals and S0 galaxies $\rho_{Fe,ICM}(z)$ 
and dividing 
this number by the comoving density of hot gas in clusters of galaxies. \\
Following the approach of Calura \& Matteucci (2003, 2004), to obtain $\rho_{Fe,ICM}(z)$  we use the present day cluster B luminosity density (LD), 
which can be calculated as:
\begin{equation}
\rho_{B, cl}(0)= \frac{\rho_{cl}}{(M/L_{B})_{cl}} = 5.4 \cdot 10^{6} L_{\odot} Mpc^{-3},
\end{equation}
where we have used the total cluster mass density 
$\rho_{cl}=1.6 \cdot 10^{9} M_{\odot} Mpc^{-3}$ (Fukugita \& Peebles 2004, hereinafter FP04, Reiprich \& B\"oringer 2002) and 
the cluster gravitational mass-to-light ratio 
$(M/L_{B})_{cl} \sim 300 M_{\odot}  L_{\odot}^{-1} $ (Fukugita, Hogan \& Peebles 1998, hereinafter FHP98, Loewenstein 2004). The 
number density of luminous galaxies in clusters can be calculated as:
\begin{equation}
n_{gal,cl}= \frac{\rho_{B, cl}}{L_{B, cl}^{*}} = 2.1 \cdot 10^{-4} Mpc^{-3},      
\end{equation}
where we have used the value $L_{B, cl}^{*} = 2.6 \cdot 10^{10} L_{\odot}$ for  the characteristic luminosity of local clusters 
(Beijersbergen et al. 2002). 
If $M_{Fe, E}(z)$ and $M_{Fe, S0}(z)$ are the cumulative masses of 
Fe ejected into the ICM at redshift $z$ by a typical elliptical and a typical S0, respectively, 
the comoving density of the Fe in the ICM is: 

\begin{equation}
\rho_{Fe, ICM} (z)=  n_{gal,cl} \cdot [f_{E} \cdot M_{Fe, E}(z) +f_{S0} \cdot M_{Fe, S0}(z)].
\end{equation}
The quantities $f_{E}$ and $f_{S0}$ are the fractions of ellipticals and S0 galaxies in clusters, respectively. 
We assume that the fraction of ellipticals does not vary with redshift. 
This is equivalent to neglect the merging of the galaxies across the redshift range studied in 
this paper. 
The validity of this approximation has been widely tested by 
Calura \& Matteucci (2003, 2006a, 2006b) and Calura, Matteucci \& Menci (2004). 
At z=0.5, we assume that a fraction  $f_{S0}$ of all the cluster galaxies starts transforming from blue discs into S0s.\\
The ICM Fe abundances observed by Balestra et al. (2007) are determined considering 
only the innermost cluster regions, i.e. within 
$r \le (0.15- 0.3)  R_{vir}$, where $R_{vir}$ is the virial radius.
To compare our predictions with the data by Balestra et al. (2007), 
we need to consider the ICM mass density within the same cluster regions as the ones explored by these authors. 
For simplicity, we assume that all the Fe ejected by luminous E and S0 galaxies in the ICM is contained within $0.3 R_{vir}$. 
This is expected on the basis of the highly concentrated 
spatial distribution of bright galaxies in clusters (Biviano \& Salucci 2006). 
\footnote{Assuming for the ICM Fe abundance profile the one found by De Grandi et al. (2004) for cool core clusters, 
which represent at least 2/3 of local clusters, we compute that the Fe mass within $0.3 R_{vir}$ is 60 \% of the total. 
However, we stress that in this paper we are mainly interested in reproducing the trend of the Fe 
abundance in the ICM with redshift, rather than the exact value of the total Fe ICM abundance. }
In the local Universe, the total ICM mass density is $\rho_{ICM}=2.45 \cdot 10^{8} M_{\odot} Mpc^{-3}$ (FP04).  
We assume that this quantity is constant across the redshift range $0\le z \le 1.5$.  
By assuming for the ICM mass density profile the $\beta$ profile $D_{ICM}(r) \propto [1 +(r/r_{c})^{2}]^{-1.5\beta}$, with 
$\beta=0.625$ and $r_{c}=0.06 R_{vir}$ (Cavaliere \& Fusco-Femiano 1978, Biviano \& Salucci 2006), 
 the fractional ICM mass within $\sim 0.3  R_{vir}$ is $ F_{0.3R_{vir}} \simeq 0.2$. 
The ICM mass density within $0.3  R_{vir}$ is simply: 
\begin{equation}
\rho_{ICM, 0.3  R_{vir}} = F_{0.3R_{vir}} \cdot \rho_{ICM} = 4.9 \cdot 10^{7} M_{\odot} Mpc^{-3}. 
\end{equation}
At this point, we can calculate the Fe abundance in the ICM 
as a function of redshift as: 
\begin{equation}
X_{Fe, ICM}(z) = \frac{\rho_{Fe, ICM}(z)}{\rho_{ICM, 0.3  R_{vir}}}.
\end{equation}
In Figure 1, we show the observed and predicted evolution of the Fe abundance in the ICM, 
calculated with the two different MR01 (left panel) and MVP06 (right panel) formulations for the Type Ia SN rate. 
In Figure 1, the observational values for the ICM Fe abundances are from Balestra et al. (2007), 
rescaled by a factor of 1.5, owing to the meteoritic value by Anders \& Grevesse (1989). 
The Fe enrichment of the ICM at redshift $z>0.5$ can be explained as due entirely to the galactic winds in ellipticals. 
We account for the  $X_{Fe,ICM}\sim 0.4 X_{Fe,\odot}$ observed at $z>0.5$ assuming for E galaxies 
a fraction of $f_{E}=0.24$ and $f_{E}=0.21$, by calculating the Ia SN rate according to the MR01 and MVP06 formalism, respectively. 
The evolution of  $X_{Fe,ICM}$ observed for $z<0.5$ allows us to determine the present-day S0 fraction,  $f_{S0}$. 
The best fit to the ICM Fe abundance and its evolution are obtained by assuming 
$f_{S0}=0.61$ and $f_{S0}=0.75$, with a Type Ia SNR calculated according to MR01 and MVP06, respectively. 
The implied total fraction of S0+E galaxies is $f_{S0+E} =0.81 - 0.96$,  
in excellent agreement with the fractions of early type galaxies observed in the central 
regions of clusters 
(Dressler 1980, Dressler et al. 1997, Andreon et al. 1997, Fasano et al. 2000, Biviano et al. 2002). 
We explain the increase of the ICM Fe abundance observed by Balestra et al. (2007) as 
due to gas ejection by the progenitors of S0 galaxies. However, we note that the study of the 
evolution of $X_{Fe,ICM}(z)$ does not allow us to determine which is the true delay time distribution for SNe Ia progenitors. 
From Figure 1, it is interesting to note that 
the Fe produced by SNe Ia in S0 galaxies for $z<0.5$ accounts only for the $\sim 10-15 \%$ of the total Fe in the ICM at $z=0$. 
This means that the majority of the Fe ejected by S0 galaxies has been produced at redshifts $z>0.5$.\\
The possible mechanisms of the galactic mass loss 
are bound to environmental effects 
and can be of various types: tidal interactions, ram pressure stripping, viscous stripping, starvation and thermal evaporation. 
Inside local clusters, 
the most efficient of all these processes  in removing the gas from large galaxies are ram pressure stripping and viscous stripping 
(Boselli \& Gavazzi 2006).  
According to our predictions, at $z=0.5$ a typical progenitor of a S0 galaxy has a gas mass of $1 \cdot 10^{10} M_{\odot}$, lost in 5 Gyr. 
For this galaxy, we predict gas ejection rates of the order of 2 $M_{\odot}/yr$. This value is in agreement with the mass loss rates predicted 
by numerical simulations. In particular, 
Schindler et al. (2005)  found that ram pressure stripping is more efficient as a gas ejection mechanism than galactic 
winds in the redshift interval between 1 and 0, with mass loss rate values between 0.4 $M_{\odot}/yr$ and $\sim 10 M_{\odot}/yr$, 
consistent with our estimates.\\

\subsection{The evolution of the Type Ia SN rate in clusters}
In Figure 2, we show the predicted evolution of the Type Ia SN rate density (SNRD) $\rho_{SNR \, Ia}$, 
expressed in $\delta^{-1} Mpc^{-3} yr^{-1}$, with $\delta$ being the cluster baryon
overdensity with respect to the field (Loewenstein 2006),  
in the case of the two Type Ia SN formulations described in section 2.1. 
In this case, we show the predicted Type Ia SN rates 
for three types of galaxies, i.e. E, S0 and spirals (Sp). 
For spiral galaxies, we use the model described in Calura \& Matteucci (2006b), i.e. the two-infall model with a Salpeter IMF. 
For E, S0 and Sp galaxies, the cluster Ia SNRD is given by: 
\begin{equation}
\rho_{SNR \, Ia, k}(z) = \rho_{B}(z)  \,  \frac{R_{SNu, k}}{100} \, yr^{-1} \frac{L_{B, \odot}}{10^{10}} \, f_{k} \, \delta^{-1}.
\label{snrdeq}
\end{equation}
(see Calura \& Matteucci 2006b), 
where $k=E, S0, Sp$ and  $ \rho_{B} (z)$ is the field B band LD. 
For the total field LD, we assume a local value of $\rho_B (0) = 1.3 \cdot 10^{8} L_{\odot} Mpc^{-3}$  
and we assume a redshift dependency $\rho_B (z)= \rho_B (0) \cdot (1+z)^{1.9}$ (Dahlen et al. 2004). 
$R_{SNu, k}$ and $f_{k}$ are the Type Ia SN rate (expressed in SNu\footnote{1 SNu = 1 SN per 100 yr per $10^{10}$ $L_{B, \odot}$.}) 
and the morphological fraction 
for the galaxy of the $k-$th morphological type, respectively. 
In Figure 2, the predicted SN rates are compared with the observations 
by Gal-Yam et al. (2002) and Sharon et al. (2006). Here, the transformation from the Type Ia SN rates 
expressed in SNu $R_{SNu}$,  
as presented in table 7 of Gal-Yam et al. (2002) and table 5 of Sharon et al. (2006), to $\rho_{SNR \, Ia}$ 
is performed according to : 
\begin{equation}
\rho_{SNR \, Ia} =  \rho_{B}  \, \frac{R_{SNu}}{100} \, yr^{-1} \, \frac{L_{B, \odot}}{10^{10}} \delta^{-1}
\end{equation}
In both panels of Fig. 2, we note that at redshift $z<1.5$, the cluster SNRD is dominated by S0 galaxies.\\ 
With both MR01 and MVP06 formulations, the predicted total cluster Ia SNRs are in very good 
agreement with the available observations. 
As can be seen in Fig. 2, the main differences in the Type Ia SNR predicted in the MR01 and MVP06 
formulations concern the earliest phases of galactic evolution, i.e. at redshifts $z> 2 $. 
In principle, observations 
of Type Ia SNR as a function of the morphological type at these redshifts could allow us to disentangle between these two formulations. 

\begin{figure*}
\leftline{\includegraphics[height=18pc,width=18pc]{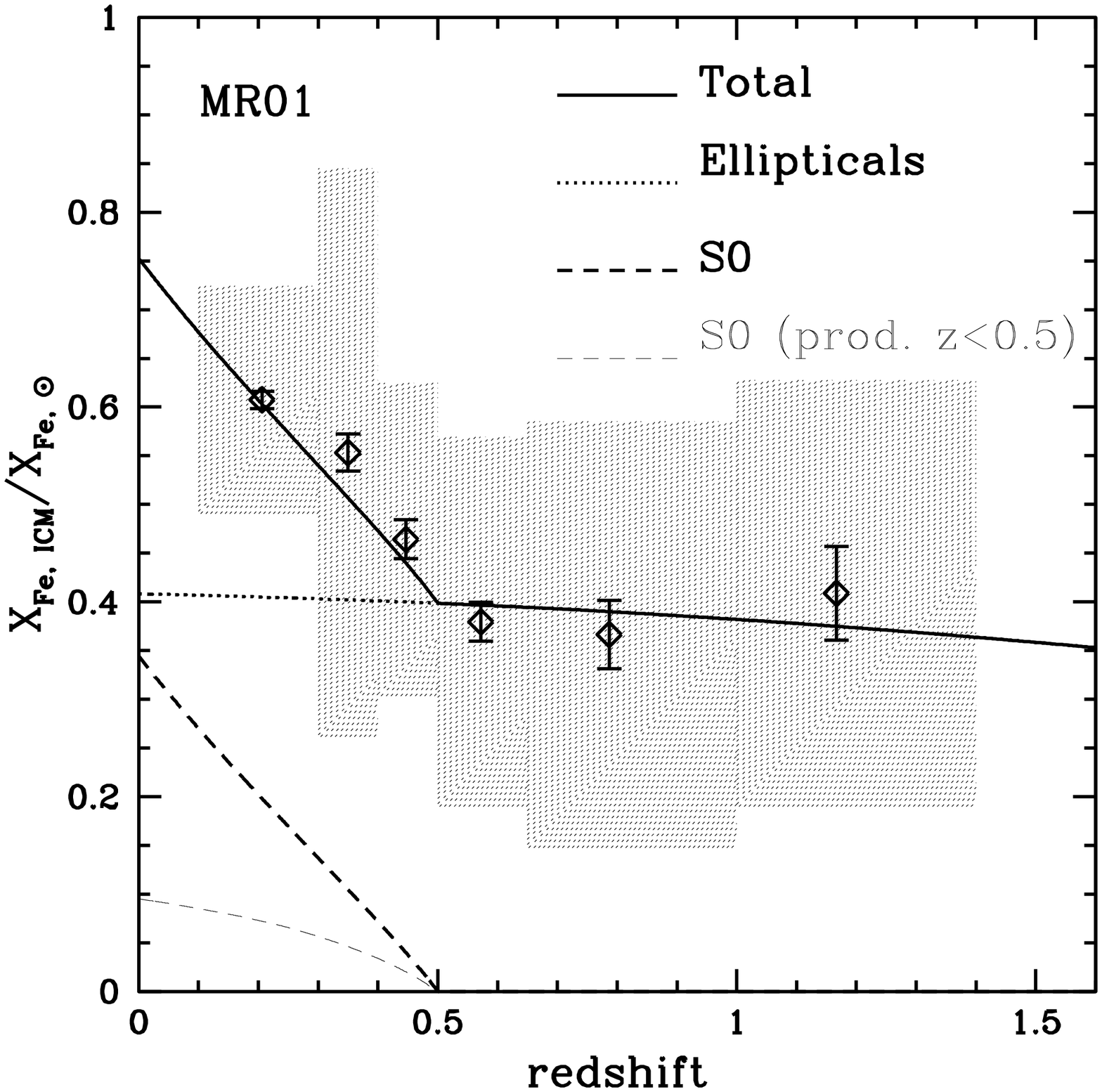}
\leftline{\includegraphics[height=18pc,width=18pc]{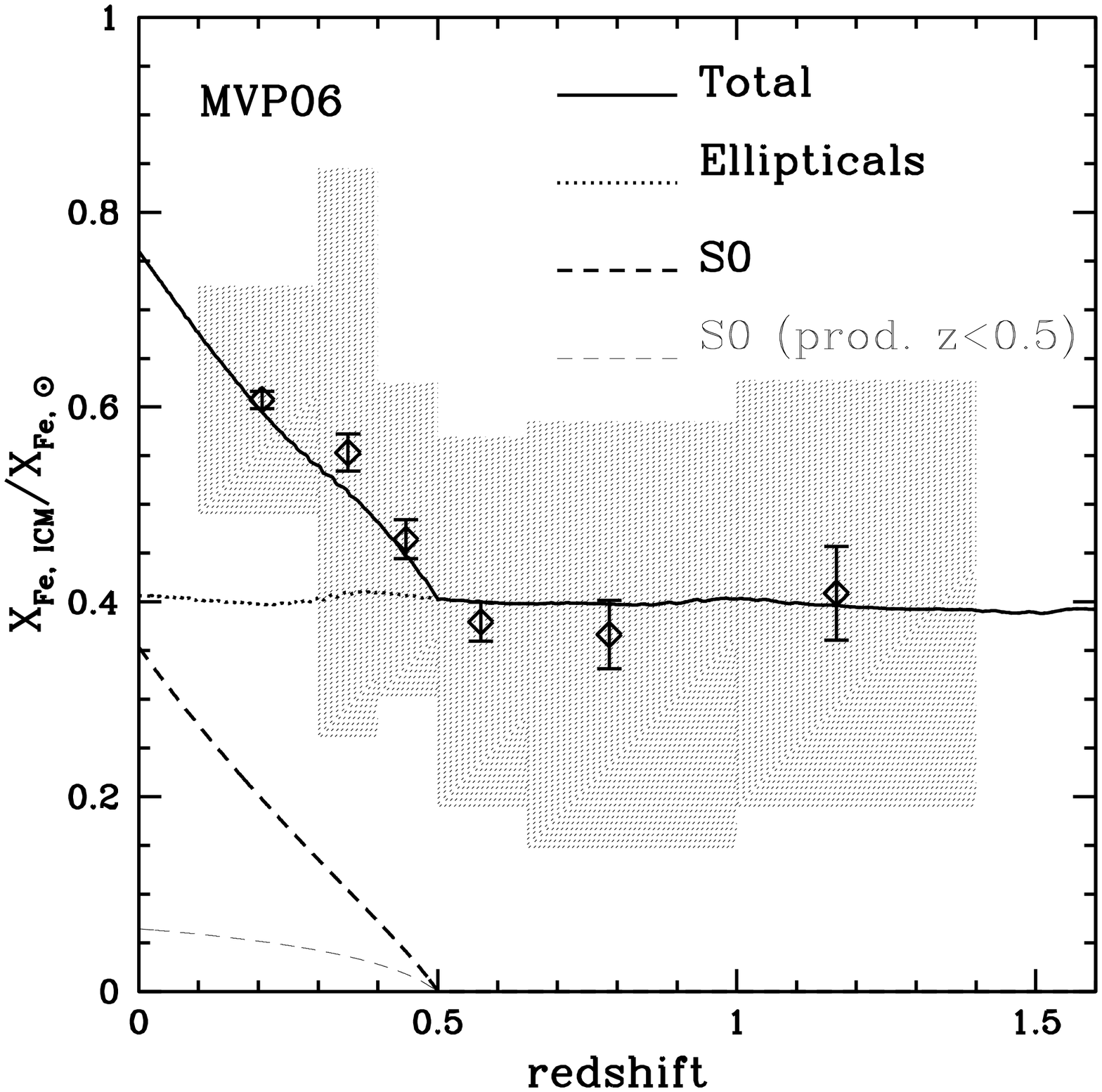}} } 
\caption[]{
Observed and predicted redshift evolution of the Fe abundance in the ICM relative to solar. 
In both panels, the open diamonds with error bars and the shaded areas show 
the  weighted means and the \emph{rms} dispersion around the weighted means of 
the Fe abundances 
in the 5 redshift bins as observed by Balestra et al. (2007), respectively. 
The Fe abundances by Balestra et al. (2007) have been  
rescaled by a factor of 1.5, owing to the meteoritic value by Anders \& Grevesse (1989). 
In the left (right) panel, the thick dashed lines, dotted lines and the solid lines are the predicted 
contributions to the ICM Fe abundance given by S0 galaxies, ellipticals and the total ICM Fe 
abundance, respectively, calculated assuming for the Ia SN rate the MR01 (MVP06) formulation. 
The thin dashed lines represent the contribution due to the Fe produced in S0 at $z<0.5$. }
\label{XFe}
\end{figure*}
\begin{figure*}
\leftline{\includegraphics[height=18pc,width=18pc]{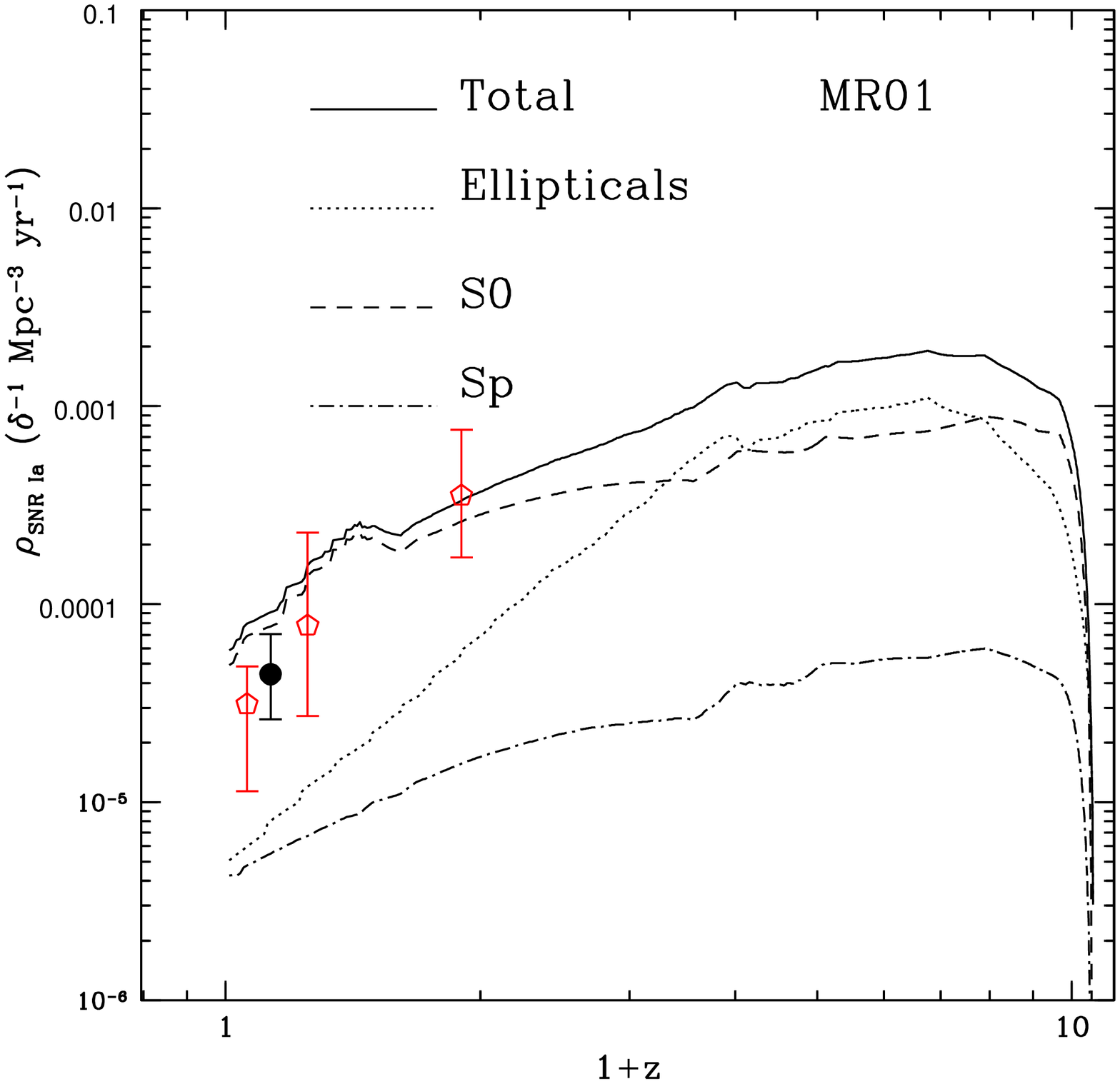}
\leftline{\includegraphics[height=18pc,width=18pc]{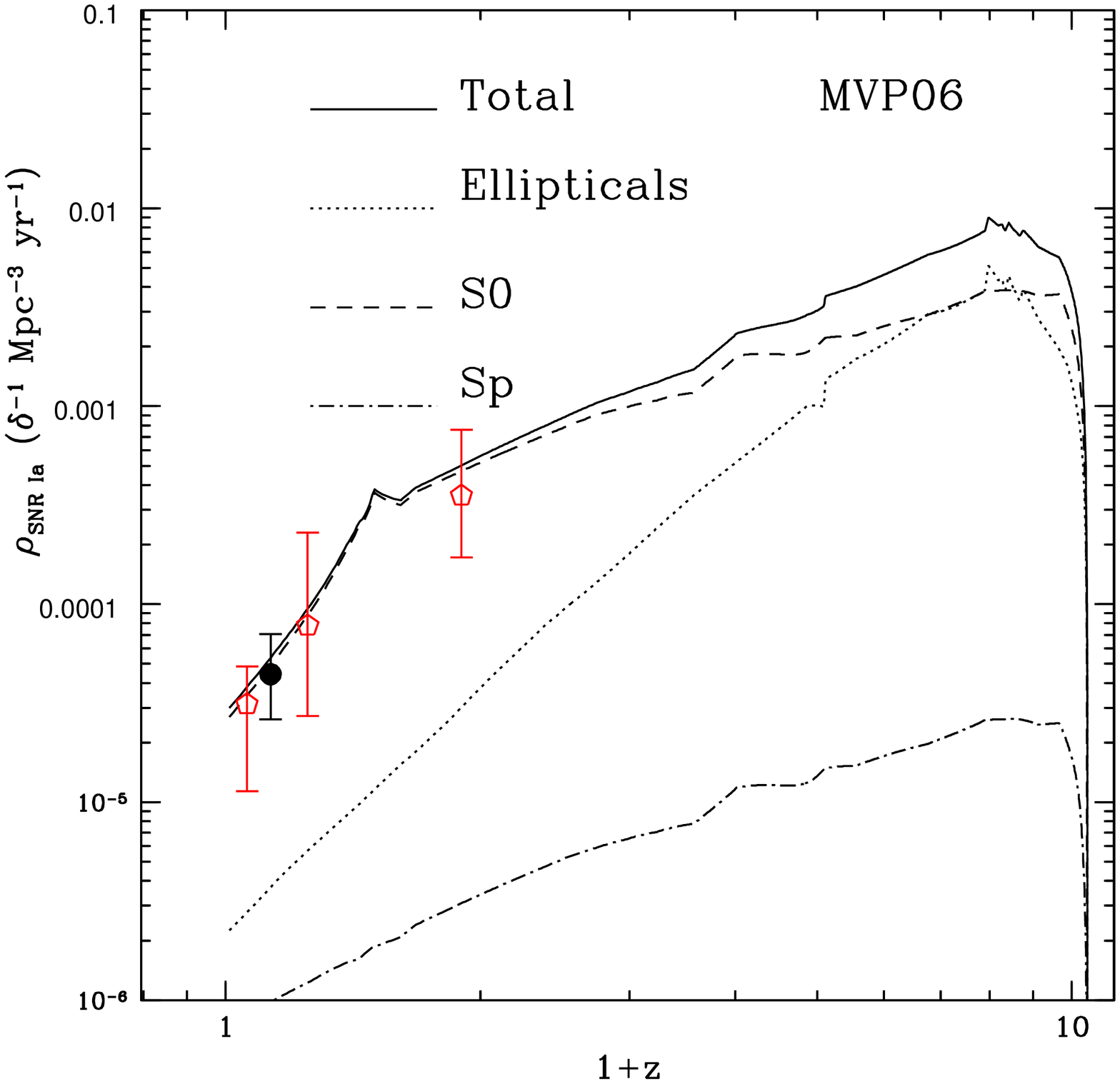}} } 
\caption[]{
Observed and predicted redshift evolution of the Type Ia SN rate density  $\rho_{SNR \, Ia}$ in clusters of galaxies, 
in units of $\delta^{-1} Mpc^{-3} yr^{-1}$.  
In both panels, the open pentagons and the solid circle are the Type Ia SN rate in clusters 
as observed by Gal-Yam et al. (2002) and Sharon et al. (2006), respectively. 
The dotted lines, dashed lines, dash-dotted and solid lines are the predicted  
cluster Type Ia SN rate for ellipticals, S0 galaxies, spirals and the total Ia SN rate, respectively.  
The Type Ia SNRs are 
calculated assuming the formulation of MR01 (left panel)  and the formulation of MVP06 (right panel).
}
\label{SNR}
\end{figure*}

\section{Conclusions}
In this Letter 
we have studied the evolution of the Fe content of clusters of galaxies, 
using chemical evolution models for galaxies of different morphological types. 
We have assumed that the Fe ICM enrichment is due to elliptical galaxies, ejecting their gas via galactic winds, 
and to S0 galaxies, which lose their ISM by means of external mechanisms, such as ram pressure stripping and viscous stripping 
(Boselli \& Gavazzi 2006). 
Our results indicate that the ICM Fe abundance observed at $z >0.5 $ can be explained thanks to the Fe-rich gas ejected by cluster 
ellipticals through SN-driven galactic winds. The increase of the ICM Fe abundance as observed by Balestra et al. (2007) at $0 \le z \le 0.5$  
can be explained by stripping of Fe-rich ISM from the progenitors of S0 galaxies in this redshift interval. 
For S0, we predict a stripping rate of $\sim 2 M_{\odot}/yr$, consistent with the ram pressure stripping rates achieved in numerical 
simulations (Schindler et al. 2005).  
The observed $X_{Fe, ICM}$ are well reproduced assuming E+S0 cluster fractions of $f_{E+S0}=0.81-0.96$,   
in excellent agreement with the available observations for the central regions of galaxy clusters. 
Beside the observed evolution of $X_{Fe, ICM}$, our picture is consistent with the observed evolution of the 
Type Ia SN rate density in clusters. 
We model the cluster Type Ia SN rate according to two different scenarios of SNe Ia progenitors, 
by adopting 
the MR01 and MVP06 formulations and 
we compare our predictions with the available observations. In both scenarios, 
the cluster Type Ia SN rate is reproduced with very good accuracy. 
However, the present data do not allow us to establish which is the best scenario for Type Ia SN progenitors. 
The observation of the SN Ia rate at redshift $z>2$ as a function of the morphological type could be helpful 
to disentangle between the two scenarios studied in this Letter.

\section*{Acknowledgments} 
FC wish to thank Antonio Pipino and Andrea Biviano 
for several interesting discussions. PT acknowledges financial contribution from contract ASI--INAF I/023/05/0
and from the PD51 INFN grant.

\label{lastpage}


\begin{thebibliography}{99}
\bibitem[]{} Anders E., Grevesse N., 1989, Geochim. Cosmochim. Acta, 53, 197			
\bibitem[]{} Asplund M., Grevesse N., Sauval A. J., 2005, in ASP Conf. Ser. 336, ``Cosmic Abundances as Records of Stellar Evolution and Nucleosynthesis'', ed. T. G. Barnes III, \& F. N. Bash (San Francisco: ASP), 25
\bibitem[]{} Arnaud M., Rothenflug R., Boulade O., Vigroux L., Vangioni-Flan E., 1992, A\&A, 254, 49	
\bibitem[]{} Andreon S., Davoust E., Poulain P., 1997, A\&AS, 126, 67
\bibitem[]{} Balestra I., Tozzi P., Ettori S., Rosati P., Borgani S., Mainieri V., Norman C., Viola M., 2007, A\&A, 462, 429
\bibitem[]{} Beijersbergen M., Hoekstra H., van Dokkum P. G., van der Hulst T., 2002, MNRAS, 329, 385						   
\bibitem[]{} Biviano A., Salucci P., 2006, A\&A, 452, 75B						   
\bibitem[]{} Biviano A., Katgert P., Thomas T., Adami C., 2002, A\&A, 387, 8B							   
\bibitem[]{} Boselli A., Gavazzi G.,  2006, PASP, 118, 517 							    
\bibitem[]{} Boselli A., Boissier S., Cortese L., Gil de Paz A., Seibert M., Madore B. F., Buat V., Martin D. C., 2006, ApJ, 651, 811B	   
\bibitem[]{} Butcher H., Oemler A. Jr., 1978, ApJ, 219, 18B	
\bibitem[]{} Calura F., Matteucci F., 2003, ApJ, 596, 734 
\bibitem[]{} Calura F., Matteucci F., 2004, MNRAS, 350, 351 
\bibitem[]{} Calura F., Matteucci F., 2006a, MNRAS, 369, 465
\bibitem[]{} Calura F., Matteucci F., 2006b, ApJ, 652, 889 						   
\bibitem[]{} Calura F., Matteucci F., Menci N., 2004, MNRAS, 353, 500  						   
\bibitem[]{} Cavaliere A., Fusco-Femiano R., 1978, A\&A, 70, 677					   
\bibitem[]{} Chiappini C., Matteucci F., Gratton R. 1997, ApJ, 477, 765						   
\bibitem[]{} Dahlen T., et al., 2004, ApJ, 613, 189								   
\bibitem[]{} Della Valle M., Panagia N., Padovani P., Cappellaro E., Mannucci F., Turatto M., 2005, ApJ, 629, 750
\bibitem[]{} Domainko W., Mair M., Kapferer W., van Kampen E., Kronberger T., Schindler S., Kimeswenger S., Ruffert M., Mangete O. E., 2006, A\&A, 452, 795					   
\bibitem[]{} Dressler A., 1980, ApJ, 236, 351					   
\bibitem[]{} Dressler A., Oemler A. Jr., Couch W. J., Smail I., Ellis R. S., Barger A., Butcher H., Poggianti B. M., Sharples R. M., 1997,  ApJ, 490, 577                                               
\bibitem[]{} Elmegreen D.  M., Elmegreen B. G., Frogel J. A., Eskridge P. B., Pogge R. W., Gallagher A., Iams J.,  2002, AJ, 124, 777
\bibitem[]{} Ettori S., 2005, MNRAS, 362, 110 									   
\bibitem[]{} Fasano G., Poggianti B. M., Couch W. J., Bettoni D., Kjaergaard P., Moles M., 2000, ApJ, 542, 673								   
\bibitem[]{} Fran\c cois P., Matteucci F., Cayrel R., Spite M., Spite F., Chiappini C., 2004, A\&A, 421, 613							   
\bibitem[]{} Fukugita M., Hogan C. J., Peebles P. J. E., 1998, ApJ, 503, 518 		
\bibitem[]{} Fukugita M., Peebles P. J. E., 2004, ApJ, 616, 643 					   
\bibitem[]{} Gal-Yam A., Maoz D., Sharon K., 2002, MNRAS, 332, 37 
\bibitem[]{} Greggio L., Renzini A., 1983, A\&A, 118, 217 							   
\bibitem[]{} Greggio L., 2005, A\&A, 441, 1055 								   
\bibitem[]{} Gunn J. E., Gott J. R., 1972, ApJ, 176, 1					   
\bibitem[]{} Helsdon S. F., Ponman T. J., 2003, MNRAS, 339, 29 	
\bibitem[]{} Iwamoto K., Brachwitz F., Nomoto K., Kishimoto N., Umeda H., Hix W. R., Thielemann, F.-K., 1999, ApJS, 125, 439I							   
\bibitem[]{} Jimenez R., MacDonald J., Dunlop J. S., Padoan P., Peacock J. A., 2004, MNRAS, 349, 240 
\bibitem[]{} Larson R. B., Tinsley B. M., Caldwell C. N., 1980, ApJ, 237, 692
\bibitem[]{} Lodders K., 2003, ApJ, 591, 1220
\bibitem[]{} Loewenstein M., 2003, in "Origin and Evolution of the Elements", eds McWilliam A., Rauch M., Carnegie Observatories Astrophysics Series, Cambridge University Press, 422  
\bibitem[]{} Loewenstein M.,  2006, ApJ, 648, 230								   
\bibitem[]{} Mannucci F., Della Valle M., Panagia N., 2006, MNRAS, 370, 773  (MVP06) 			   
\bibitem[]{} Matteucci F., Greggio L., 1986, A\&A, 154, 279							   
\bibitem[]{} Matteucci F., Vettolani G., 1988, A\&A, 202, 21
\bibitem[]{} Matteucci F., Recchi S., 2001, ApJ, 558, 351   (MR01)						   
\bibitem[]{} Matteucci F., Renda A., Pipino A., Della Valle M., 2003, A\&A, 405, 23 							   
\bibitem[]{} Matteucci F., Panagia N., Pipino A., Mannucci F., Recchi S., Della Valle, M., 2006, MNRAS, 372, 265 (M06)					   
\bibitem[]{} Pipino A., Matteucci F., Borgani S., Biviano A., 2002, NewA, 7, 227									   
\bibitem[]{} Poggianti B. M., Smail I., Dressler A., Couch W. J., Barger A. J., Butcher H., Ellis R. S., Oemler A. Jr., 1999, ApJ, 518, 576							   
\bibitem[]{} Reiprich T .H., B\"ohringer H., 2002, ApJ, 567, 716					  
\bibitem[]{} Renzini A., Ciotti L., D'Ercole A., Pellegrini S., 1993, ApJ, 419, 52
\bibitem[]{} Renzini A., 1997, ApJ, 488, 35				   
\bibitem[]{} Romano D., Matteucci F., Salucci P., Chiappini C., 2000, ApJ, 539, 235
\bibitem[]{} Salpeter E. E., 1955, ApJ, 121, 161							   
\bibitem[]{} Schweizer F., Seitzer P., 1992, AJ, 104, 1039
\bibitem[]{} Sharon K., Gal-Yam A., Maoz D., Filippenko A. V., GuhaThakurta P., 2006, ApJ, submitted, astro-ph/0610228						   
\bibitem[]{} Schindler S., Kapferer W., Domainko W., Mair M., van Kampen E., Kronberger T., Kimeswenger S., Ruffert M., Mangete O., Breitschwerdt D.,  2005, A\&A, 435, 25				
\bibitem[]{} Tozzi P., Rosati P., Ettori S., Borgani S., Mainieri V., Norman C., 2003, ApJ, 593 705								   
\bibitem[]{} Tran K.-V. H., van Dokkum P., Illingworth G. D., Kelson D., Gonzalez A., Franx M., 2005, ApJ, 619, 134							   
\bibitem[]{} van den Bergh, S. 1976, ApJ, 206, 883								   
\bibitem[]{} Whelan J., Iben I. Jr., 1973, ApJ, 186, 1007	
\bibitem[]{} Woosley S.E., Weaver T.A., 1995, ApJS, 101, 181
\bibitem[]{} van Dokkum P. G., Franx M., Fabricant D., Illingworth G. D., Kelson D. D., 2000, ApJ, 541, 95

\end{thebibliography}
\end{document}